# Effects of surface compliance and relaxation on the frictional properties of lamellar materials

Alex Smolyanitsky[1*], Shuze Zhu[2], Zhao Deng[1,3], Teng Li[2,3] and Rachel J. Cannara[1§]

[1]Center for Nanoscale Science and Technology, National Institute of Standards and Technology, Gaithersburg, MD 20899

[2]Department of Mechanical Engineering, University of Maryland, College Park, Maryland 20742

[3]Maryland NanoCenter, University of Maryland, College Park, MD 20742

[*]To whom correspondence should be addressed: alex.smolyanitsky@nist.gov

[§] Present address: Intellectual Ventures Laboratory, Bellevue, WA 98005

**Abstract**

We describe the results of atomic-level stick-slip friction measurements performed on chemically-modified graphite, using atomic force microscopy (AFM). Through detailed molecular dynamics simulations, coarse-grained simulations, and theoretical arguments, we report on complex indentation profiles during AFM scans involving local reversible exfoliation of the top layer of graphene from the underlying graphite sample and its effect on the measured friction force during retraction of the scanning tip. In particular, we report nearly constant lateral stick-slip magnitudes at decreasing loads, which cannot be explained within the standard framework based on continuum mechanics models for the contact area. We explain this anomalous behavior by introducing the effect of local compliance of the topmost graphene layer, which varies when interaction with the AFM tip is enhanced. Such behavior is not observed for non-lamellar materials. We extend our discussion toward the more general understanding of the effects of the top layer relaxation on the friction force under positive and negative loads. Our results may provide a more comprehensive understanding of the effectively negative coefficient of friction recently observed on chemically-modified graphite.

**Introduction**

Frictional properties of atomically thin layers and lamellar materials at the nanoscale have become a research topic of great interest. The investigative effort is fueled by the promise of a novel class of solid-state lubricants, as well as several seeming anomalies observed experimentally and predicted theoretically. Recently, atomic force microscopy (AFM) studies of friction in substrate-bound monolayer and bilayer graphene revealed unusually low friction attributed in part to electron-phonon scattering specific to graphene [1]. This finding suggested the use of graphene as an atomically thin lubricant in nanomechanical systems. Further, AFM studies of friction in stacked free-standing and substrate-bound atomic layers showed, in addition to low friction, a dependence of the kinetic friction coefficient on the number of stacked layers [2]. Asymmetric "puckering" of layers in front of the scanning AFM tip was suggested as responsible for this behavior [2]. A theoretical follow-up attributed the observed behavior, at least for tips of a few nanometers in diameter, to a non-destructive form of viscoelastic ploughing [3], which can be viewed as a special case of a general ploughing process [4, 5]. The latter work laid the basis for further theoretical investigation, which revealed a possibility for free-standing monolayer and few-layer graphene to exhibit anomalous frictional behavior, whereby friction *increases* as the AFM tip is further retracted at negative normal loads, resulting in a locally negative slope of the friction-load curve [6].

Concurrently, in a recent work by Z. Deng and colleagues, AFM studies of friction on chemically-modified graphite revealed that a similar negative friction coefficient is possible at both positive (pushing) and negative (pulling) normal loads under retraction. This behavior was attributed to the interplay between the tip-sample interactions and the interlayer interactions, resulting in complex contact deformations in lamellar (or layered) samples [7, 8]. A number of interesting friction-load properties of suspended and substrate-bound few-layer graphene were found through further experimental work [9]. However, this work revealed an initial negative friction-load slope, followed by positive slope under negative loads, in seeming contradiction with the theoretical predictions. Despite the research effort, the complexity of physical mechanisms responsible for the observed frictional properties in stand-alone and stacked atomically thin layers is far from well understood. Moreover, the established continuum mechanical models based on Hertzian contact theory [10], from Johnson, Kendall, and Roberts (JKR) [11] to Derjaguin, Muller, and Toporov (DMT) [12], appear to be inadequate for describing both the experimental and simulated findings.

In this work, we further elucidate the physical processes responsible for the effectively negative friction coefficient recently reported for chemically-modified graphite surfaces [7] and aged molybdenum disulfide (see Supplementary Information (SI)). Analysis of the experimentally observed lateral stick-slip [13] experienced by an AFM tip, combined with molecular dynamics (MD) simulations, coarse-grained (CG) simulations and theoretical estimates, reveal the complex dynamic relaxation of the uppermost graphite layers around the AFM tip. We attribute the observed behavior to the enhanced interactions of the tip with the sample, compared with the graphite interlayer interactions. We demonstrate that, depending on the size of the tip, these interactions result in local deformations of the layers in contact with the tip, leading to a combination of contributions to friction both by conventional van der Waals bonding and debonding and by the aforementioned viscoelastic ploughing. We hypothesize that the fine interplay between these contributions can indeed result in a number of interesting frictional properties of lamellar materials at the nanoscale, including the observed negative friction coefficient.

**Methods and systems**

The experimental data consist of friction loops recorded at individual loads for a 30 nm radius silicon nitride AFM tip scanning on freshly-cleaved versus aged highly-oriented pyrolytic graphite samples, as described elsewhere [7]. In Ref. [7], friction forces as a function of applied normal load were calculated in the standard way by taking half the average difference between lateral forces measured for forward and backward line scans at a given load value. The thermal noise method [14] and a diamagnetic lateral force calibrator [15] were used to calibrate normal and lateral forces, respectively. Stick-slip data were automatically recorded within individual line scans, thus enabling us to analyze detailed information about the contact behavior obtained concurrently with the friction-load data reported in Ref. [7]. The measurement error in the friction-load plots is comparable to the size of the data points, and data scatter is due to small inhomogeneities of the sample, as each data point was measured on a slightly different scan line.

The MD simulation setup within the framework used previously in [3, 6, 7] may be found in Ref. [7]. The simulated graphite sample consisted of five 5.5 nm × 6.2 nm layers of graphene stacked in an AB order. The interatomic interactions within each graphene layer, as well as the tip were determined by the well-established Tersoff-Brenner potential [16, 17]. The interlayer interactions, and the tip-sample interactions were defined by the Lennard-Jones potential [18] with $\varepsilon = 7.5$ meV and $\sigma = 0.31$ nm, resulting in an interlayer adhesion of 42.8 meV/atom, in agreement with *ab initio* and experimental data [19-21]. All MD simulations were carried out at 300 K, maintained by the Langevin thermostat [22], as parameterized elsewhere [3]. The atoms at the edge of all layers in the X-direction and all atoms in the lowest layer were harmonically restrained against displacement, representing a fixed boundary. A periodic boundary was imposed in the Y-direction. A flat, rigid 2.5 nm × 1.5 nm graphene plate was used to model the tip scanning along the Y-direction. We employed a flat plate instead of a round sphere for these atomistic simulations to eliminate variations in friction forces and stick-slip amplitudes related simply to changes in contact area (tip geometry). During the 1.5 ns long scan, the tip was retracted away from the sample at a constant rate of 0.2 m/s, while translating laterally at 5 m/s. The initial tip-sample configuration corresponded to a normal contact force of +12 nN calculated for the tip-sample interatomic interaction strength equal to that between the stacked layers.

Using a CG simulation scheme developed by Zhu *et al*. [23] (also, see SI) , we simulated the graphite sample with four stacked layers of graphene. Although a slightly thicker stack of graphene layers would have been desirable, computational limitations at this size scale restricted the number of layers. Each graphene layer was coarse-grained; however, the AFM tip was modeled by a CG monolayer of graphene with a fixed parabolic geometry. The radius of curvature at the trough of the parabola was 25 nm. The tip was constrained to move in the direction normal to the sample surface. The dimensions of each graphene layer in the sample were 100 nm × 100 nm, as shown in Fig. S1 (b) of SI. The CG beads at the edge of all four graphene layers were fixed. We considered various loads ranging from -50 nN to +50 nN. The initial tip-sample configuration corresponded to a load of +100 nN, followed by a ramp to a desired load within the range mentioned above. The AFM tip position was modulated using an applied force with viscous damping, to expedite equilibration at each load value. The CG simulations were performed on a canonical ensemble at a constant temperature of 300 K, maintained by a Nosé-Hoover thermostat [24].

All CG simulations were carried out using the Large-scale Atomic/Molecular Massively Parallel Simulator (LAMMPS) [25]. Two sets of tip-sample adhesion were considered: (*i*) a tip-sample interaction strength equal to that between the graphene layers of the sample, and (*ii*) a three-fold enhanced tip-sample interaction as compared with the interlayer interactions. Two limiting boundary conditions were considered: (*i*) a supported case, whereby the bottom layer of graphene was restrained against deformation, and (*ii*) a suspended case, with the bottom layer free. The contact area between the tip and the top layer, as well as the local interlayer delamination area, was quantified by counting the number of CG beads that were separated by a specific distance beyond the equilibrium separation. A threshold value of 0.425 nm was used, which is greater than the equilibrium interlayer spacing in graphite by approximately 25 %.

**Results and discussion**

The stick-slip phenomenon resulting in periodic lateral motion of the AFM probe [13] at an atomic scale provides key information for understanding friction at the nanoscale. In the case of constant contact area, the stick-slip amplitude is a measure of the strength of interaction between the tip and the sample. Within existing tribological models, the interaction strength can be viewed as generally proportional to the tip-sample average contact area until reaching saturation at high normal loads [11, 12]. Therefore, a way to determine the qualitative relationship between the contact area and the normal load is to find a statistically significant relationship between the experimentally obtained stick-slip magnitude throughout a lateral scan trace at a given normal load. Provided that each trace in the X-direction (Fig. 1) is performed at a constant load, we sought a per-trace figure of merit describing the average stick-slip magnitude throughout the trace.

Because the shape of the stick-slip signal is roughly similar to a sinusoid, such a number may be estimated via Fourier transformation. Defining spectral power as the variation in lateral force signal $F(x)$, as shown at the center of Fig. 1:

$$g(k_x) = \int_0^{x_{max}} (F(x) - \bar{F}) e^{-ik_x x} dx, \qquad (1)$$

where $k_x$ is the magnitude of the wave-vector along the direction of the trace, $x_{max}$ is the scanning range along the X-axis, and $\bar{F}$ is the stick-slip average obtained from the trace to ensure that no offset from zero is present. We then define an effective stick-slip magnitude variation "energy" corresponding to the $\lambda = 0.25$ nm (2.5 Å) periodicity, the lattice constant of graphene:

$$E(k_0) = \int_{k_0 - \delta}^{k_0 + \delta} |g(k_x)|^2 dk_x, \qquad (2)$$

where $k_0 = \frac{2\pi}{\lambda} = 25.1$ nm$^{-1}$, and $\delta$ is a wave-vector range parameter, set to account for the experimentally observed variation of the stick-slip periodicity due to the thermal motion of atoms, added measurement noise, and any spurious AFM tip motion. A value of $\delta = 6.3$ nm$^{-1}$ was used, as explained in SI. The calculation described above is mathematically equivalent to applying a band-pass filter (with bandwidth set by $\delta$) to the raw trace data, followed by calculating the effective square deviation.

Given the measure provided by Eq. (2), the entire stick-slip trace along the X-direction was statistically characterized by a single number, as a function of the normal load maintained constant along that trace. It is worth noting that the values of $E(k_0)$ possess natural periodicity between traces, associated with the periodicity of tip-sample interactions in the Y-direction. We therefore average the $E(k_0)$ values within a "stack" of X-traces within $0.25\ nm$ long regions in the Y-direction, *i.e.* reporting a single averaged value every $0.25\ nm$ in Y-direction for further clarity. Finally, in the data presented here, we converted the length units in the Y-direction of the trace image to the scanning time. The calculated values of $E(k_0)$ are reported only for traces during which the tip is in contact with the sample.

Shown in Fig. 2 (a) are the stick-slip magnitude values, calculated as described above, during the AFM tip approach and retraction, while the tip is in contact with the surface of a sample and scanning across it. For the freshly-cleaved sample, the stick-slip variation (and thus the effective contact area) increases during approach and decreases during retraction, as qualitatively predicted by existing contact models [11, 12]. For aged samples, which have chemically-modified surfaces, the behavior is remarkably different. The stick-slip magnitude remains nearly constant during retraction, which suggests a constant effective tip-sample contact area, despite the ever-decreasing load. We attribute this behavior to a different pattern in the tip-sample deformation region, as discussed below. We considered the possibility of multiple stick-slip [26] as an additional contribution to the observed dependence of the stick-slip magnitude on applied load. However, provided that multiple stick-slip is likely to occur due to increased lateral stiffness of the AFM cantilever, accompanied by considerable modification of the stick-slip periodicity, a significant contribution in this case is doubtful.

Fig. 2 (b) shows the friction-load curves for the freshly-cleaved and aged samples. As reported previously, the overall magnitude of friction for the freshly-cleaved samples is low, and approaches the noise limits of the apparatus [7]. The combined linear fit to the data obtained for three freshly-cleaved datasets yields a friction coefficient of ≈ 0.0001, in general agreement with literature [1, 2, 27]. For the aged samples, however, our data indicate a clear inverse dependence of friction on the load, resulting in negative friction coefficient values, as shown in Fig. 2 (b), and also reported previously [7]. Moreover, Fig. 2 (b) shows that the AFM tip pull-off force (determined by the leftmost load values) occurs at loads approximately three times larger in magnitude on aged samples as compared with freshly-cleaved samples. According to existing contact models that incorporate interfacial adhesion [11, 12], this suggests a roughly trifold tip-sample adhesion in the aged samples, relative to the fresh samples. The variation in pull-off forces is characteristic of the typical data scatter observed for aged samples.

We observe that the inverse friction-load dependence takes place while the contact area is effectively insensitive to decreasing load. Within existing tribological models, where variations in friction are exclusively a result of changing contact area between the tip and the sample (in the absence of wear and material displacement), this finding suggests possible additional mechanisms of frictional loss in our system. In order to consider in greater detail the qualitative observation that the stick-slip magnitude is insensitive to the process of retraction, we performed atomistic simulations of the retraction process, using MD. Shown in Fig. 3a are the variations in lateral and normal forces as a function of time during retraction of a flat plate (mimicking a small region of a spherical tip of large diameter), while laterally scanning over a graphite surface. In the absence of enhancement, the lateral trace amplitude decreases continuously, followed by the gradual reduction in tip-sample contact area (until pull-off), as expected. However, in the case of an enhanced tip-sample interaction, as shown in Fig. 3 (b), there exists a region of nearly constant stick-slip magnitude that is qualitatively similar to the experimental observations in Fig. 2 (a). The effective normal load remains constant during retraction, as the top layer adheres to the retracting tip for a considerable distance along the normal direction, resulting in a local temporary delamination from the subsurface layers.

Although we observed qualitative similarity between Figs. 2 (a) and 3 (b), there is a key difference in the load regimes. An inverse friction-load dependence under retraction has been demonstrated previously [7]. These same experimental data, analyzed here, also exhibit this anomalous frictional behavior at positive loads, for which local lifting of layers would seem unlikely and was not observed in previous MD simulations using small, round scanning tips [7]. We believe that, on a larger spatial scale, a more complex form of local lifting of the top layer takes place under positive normal loads, which leads to additional frictional mechanisms compared to established models in nanotribology. Shown in Fig. 4 (a) is a typical cross-section of the tip-sample contact in a CG simulation of the supported graphite system. The form of local lifting at the outer region of the crater-like tip-sample contact presented in Fig. 4 (a) may clarify the nature of the additional sources of friction suspected to contribute to our experimental observations.

Further parameterization of existing contact models may implicitly account for a crater-like contact, if the elastic properties were forced to vary with both the adhesive force and applied load. However, a less speculative discussion is possible. Instead, the observed behavior may be related to the fundamental assumption that friction is proportional to the contact area [10-12] plus a ploughing component, when appropriate [5]. Here, we account for all of the adhesion- and load-dependent areas that can develop in the contact zone, relaxing the requirement that the relevant area be in physical contact with the tip. The total friction force $F_{tot}$ may then be expressed as a sum of contributing mechanisms, as is typical, but for a more complex contact geometry than that considered previously [5, 6]:

$$F_{tot} = \tau A_{contact} + \tau_s A_{delam} + F_p(h), \qquad (3)$$

where $\tau$ is the shear strength for the tip-sample interface and $\tau_s$ is the effective energy loss factor associated with van der Waals bond deformation in the subsurface volume during sliding. $A_{contact}$ and $A_{delam}$ are the physical contact area and the delaminated area (the effective cross-section of the subsurface volume mentioned above), respectively. $F_p(h)$ is the (non-destructive) viscoelastic ploughing contribution, which increases with the height $h$ of the lifted surface asperity [5, 6], as described in Fig. 4 (a). With this model, the effect of increasing friction with decreasing load emerges. The contact area ($A_{contact}$) may not change significantly during retraction due to strong adhesive contact between the flexible top layer and the tip. This results in a nearly constant $\tau A_{contact}$, as observed for the aged samples in Fig. 2 (a). Meanwhile, both $\tau_s A_{delam}$ and $F_p(h)$ may *increase* on retraction, resulting in an increase in the overall friction force.

We performed CG simulations to better simulate the spatial scale of the experimental system and test the argument presented above. Figs. 4 (b) and (c) show the dependence of $A_{contact}$ on the applied load for two extreme cases, supported and suspended, to separate the results from the effect of bulk stiffness. The tip-sample interaction was either equal to the interlayer interaction or enhanced by a factor of three. As shown in Fig. 4 (b, c), in the absence of enhancement, the contact area decreases with decreasing load, in agreement with established contact models [11, 12] and the experimental data for freshly-cleaved samples (Fig. 2 (a)). We also observed a saturation trend toward higher positive loads, as expected for the supported simulated system, which is significantly stiffer than graphite. When the tip-sample adhesion is enhanced, however, the load dependence is qualitatively different, as shown in Fig. 4 (b, c). For the supported limiting case, the overall contact area is insensitive to load, in agreement with the experimental data in Fig. 2 (a) for the aged samples. In the suspended limiting case, $A_{contact}$ is also insensitive to positive loads, yet readily decreases in the negative load regime. This difference may be explained by the high deformability of the suspended sample, as compared with the experimental system.

The load dependence of $A_{delam}$ and $h$ (defined schematically in Fig. 4 (a)) for the supported graphite model is shown in Fig. 5. Only the supported case is presented, because in the suspended case all four layers tend to conform to the tip, eliminating the effect of local delamination. For the case of tip-sample adhesion equal to that between layers, as shown in Fig. 5 (a), there appears to be a quick saturation of both $A_{delam}$ and $h$, which remain nearly constant at loads above +10 nN. Note that $A_{delam}$ and $h$ are low, as expected, because ideally the effect of local delamination in the absence of tip-sample interaction enhancement should be minimal. The situation changes considerably when the tip-sample interaction is enhanced by a factor of three, as shown in Fig. 5 (b). The base values of $A_{delam}$ and $h$ are significantly higher, and both quantities increase with decreasing load. Assuming that the actual system is closer to the supported case, an approximate estimate of the friction coefficient at positive loads is now possible, based on the data in Figs. 2, 4, and 5. Assuming $\tau$ and $\tau_s$ are load-independent, the friction coefficient may be calculated from Eq. (3) as

$$\alpha = \frac{dF_{tot}}{dL} = \tau \left( \frac{dA_{contact}}{dL} + \frac{\tau_s}{\tau} \frac{dA_{delam}}{dL} \right) + \frac{dF_p(h)}{dh} \frac{dh}{dL}, \qquad (4)$$

where $L$ is the normal load. Given that $\tau$ and $\tau_s$ correspond to complete bonding-debonding and mostly bond deformation, respectively, we expect $0 < \frac{\tau_s}{\tau} < 1$. At positive loads, with an "enhanced" value of $\tau \approx 300\ MPa$ [7], $\frac{dA_{contact}}{dL} \approx 0$ (Fig. 2 (a) and Fig. 4 (c)), and $\frac{dA_{delam}}{dL} \approx -0.05\ \frac{nm^2}{nN}$ (Fig. 5 (b)), it is possible to estimate $\alpha$ for a tip of known diameter. With $\frac{\tau_s}{\tau} = \frac{1}{3}$ (assumed to be roughly equal to the ratio between the strengths of tip-sample adhesion and top layer to the subsurface layer adhesion), and neglecting the ploughing contribution for the moment, we obtain $\alpha \approx -0.005$. With the contribution from the subsurface interactions reduced even further $\left( \frac{\tau_s}{\tau} = \frac{1}{6} \right)$, we obtain $\alpha \approx -0.0025$. Therefore, even without the contribution from viscoelastic ploughing, the calculated values of $\alpha$ are comparable to those shown in Fig. 2 (b).

The effect of $F_p(h)$ may be considerable, further enhancing the observed effect. We note that changes in *h* with varying positive load are not particularly significant, as shown in Fig. 5 (b). However, we recall that $F_p(h)$ is highly nonlinear with respect to the value of *h* [6], and thus the contribution from $\frac{dF_p(h)}{dh}\frac{dh}{dL}$ in Eq. (4) may not be negligible overall.

Given the above results, local lifting at the outer circular region of the contact is not only possible, but also suggests an increase of $\tau_s A_{delam}$ and $F_p(h)$ terms in Eq. (3) with decreasing *positive* load. In other words, a crater-like deformation of the top layer is present when the strength of the tip's interaction with the top layer significantly exceeds the interlayer interaction strength. The height of the rim of the crater at a given load determines the effective height of the asperity surrounding and ahead of the tip during scanning, and depends primarily on two physical properties of the lamellar system: (*i*) the amount of tip-sample adhesion enhancement, and (*ii*) the bending rigidity of a single graphene layer, which cannot be neglected [13, 28, 29]. The presence of flexible atomically-thin layers with a finite ability to bend around the tip changes the nature of the tip-sample contact. In our case, the AFM tip is scanning a locally and anisotropically soft material capable of significant and complex deformation near the tip, while remaining relatively close to a conventional solid only a few atomic layers below. This observation provides immediate insight into the reason why this effect is not observed even in free-standing graphene for round AFM tips of a few nanometers in diameter [3, 6, 7]. The energy associated with bending increases significantly with decreasing tip radius, making the crater-like deformation energetically unfavorable. With the AFM tips of ≈ 30 nm to 50 nm in diameter, as confirmed by our CG simulations, the described deformation readily takes place at the contact.

Based on the presence of a circular asperity around the tip apex, we can now explain the behavior observed experimentally. The locally lifted circular region around the tip acts as a material barrier, or wall, with viscoelastic return, which introduces a ploughing component to friction in addition to the "standard" van der Waals bonding-debonding contribution. In addition, it creates a locally perturbed van der Waals region between the top layer and the first subsurface layer, which contributes an additional van der Waals bonding-debonding frictional component. Our CG simulations suggest that the asperity height, as well as the locally delaminated area increase with decreasing positive load. Therefore, the overall amount of friction can increase with decreasing load, as long as the tip radius and the bending rigidity of the stacked layers are such that these contributions are significant.

In the experiments, chemical modification of the surface played a critical role, resulting in a significant enhancement of the tip to top layer interaction. For macroscopic tip radii, the observed anomalous friction effect is expected to cease, because the first term in Eq. (3) will eventually become the dominating contribution. It is noteworthy that the chemistry of surface oxidation and the structure of graphene oxide [30] may result in a form of subsurface oxygen intercalation, leading to local lifting of not one, but of several graphene layers. Indeed, we have observed some indication of interlayer stratification when aging graphite surfaces under ultraviolet light, but more work would be required to make conclusive statements regarding the presence and effect of intercalants. It is also possible that the considerably longer-range electrostatic tip-sample interactions arising from even minor tribocharging [31] can contribute to the local lifting of the surface layers, further enhancing the described effect. The mechanism described here explains why this behavior is highly unlikely to occur in isotropic solids, but may remain possible for soft, layered polymeric material capable of significant deformation around the tip under positive or negative loads [32].

**Conclusion**

We have presented experimental and theoretical considerations explaining the phenomenon of increasing friction under decreasing normal load observed in AFM scans of mildly oxidized graphite surfaces. We find that the lateral stick-slip magnitude remains relatively constant during tip retraction, which suggests a tip-sample contact considerably stronger than that described by existing tribological models for isotropic solids. The results of our experimental measurements and numerical simulations suggest that a complex crater-like deformation of the top layer (or the few near-surface layers) forms around the tip. Such a deformation is possible when the tip-surface interaction strength is considerably higher than the interlayer interactions. The crater-like deformation results in two additional contributions to friction. The first is an additional van der Waals bonding-debonding component between the top layer and the subsurface layer. The second is a raised, ring-like asperity at the outer circular region of the contact. These contributions to friction can be large enough to result in the experimentally observed frictional behavior.

We further demonstrate that local relaxation of the top layer around the AFM tip depends on the amount of tip-top layer interaction enhancement, the AFM tip radius, as well as the bending rigidity of a layer in a lamellar stack. As a result, the type of material probed by the AFM here is that of a locally soft, atomically-thin material capable of significant deformation around the tip, while remaining close to an anisotropic solid only a few atomic layers below the surface.

**Acknowledgement**

TL and SZ acknowledge the support by the National Science Foundation (Grant Numbers: 1069076 and 1129826). SZ thanks the support of the Clark School Future Faculty Program at the University of Maryland.

**Figures**

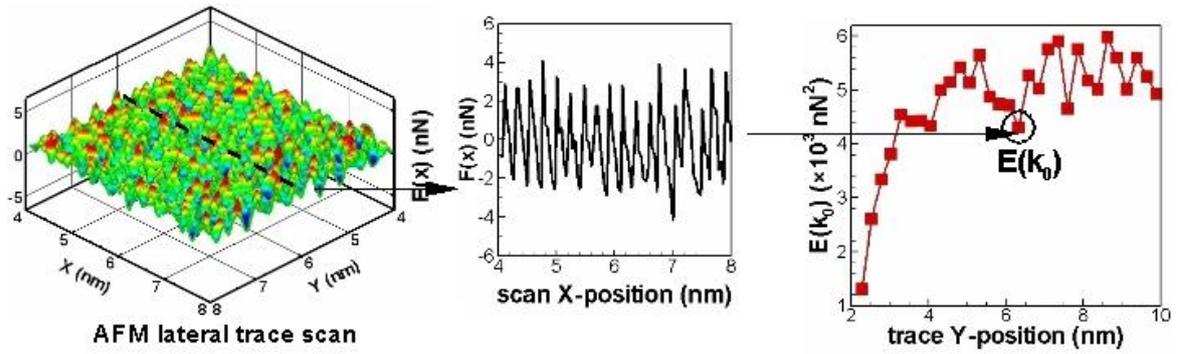

Figure 1. Calculation of the stick-slip spectral energy for a single trace from a two-dimensional lateral trace AFM scan.

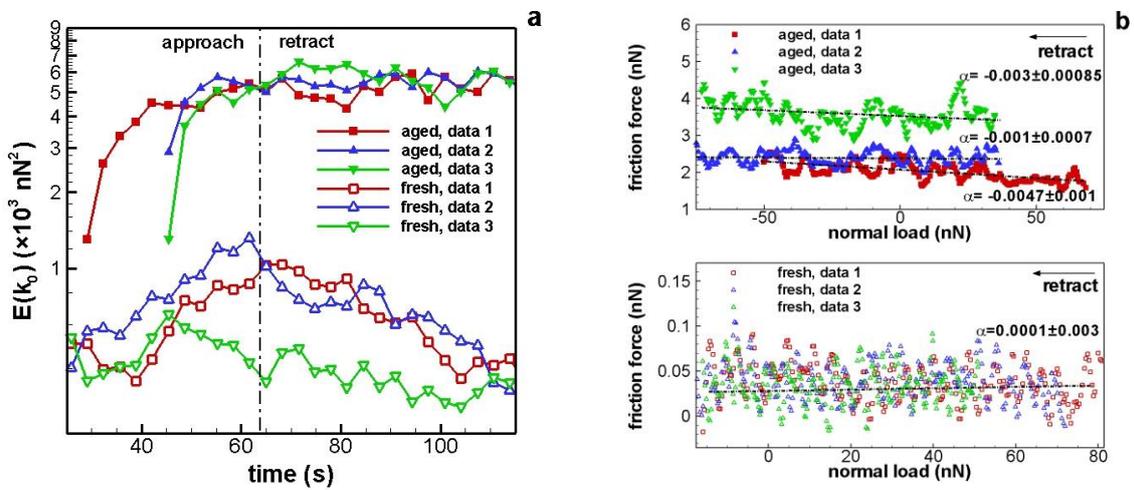

Figure 2. Spectral energies of stick-slip variation as a function of scan time (a) and friction as a function of load during retraction (b); dashed lines represent linear fits to corresponding data. The slopes and their single standard deviation uncertainties are listed to the right of each curve in (b). For a discussion of these uncertainties, see Ref. [7]. Different datasets correspond to different scanned regions of the fresh and aged samples.

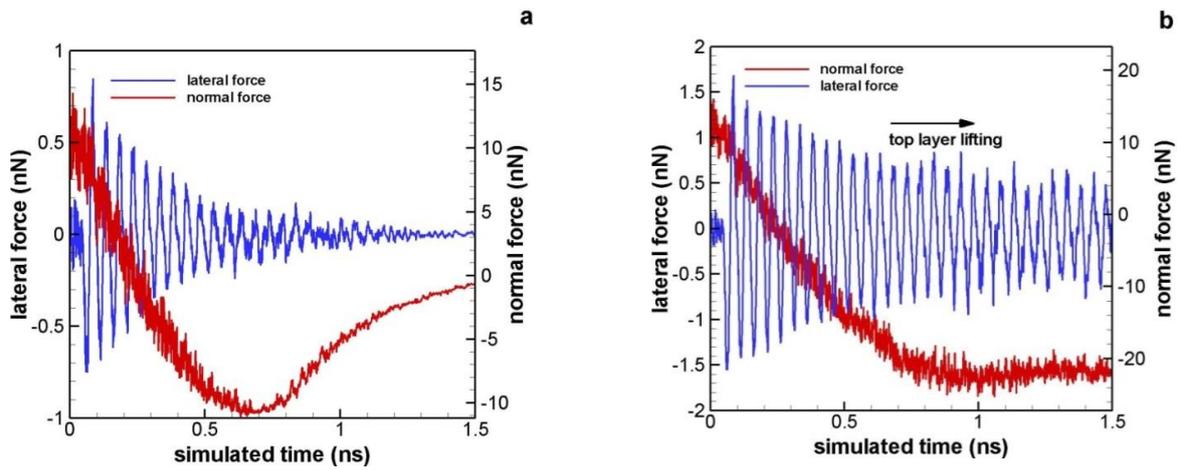

Figure 3. Simulated stick-slip and normal loads during dynamic retraction for (a) a tip-sample interaction strength equal to that of the interlayer interaction, and (b) a tip-sample interaction strength increased by a factor of three.

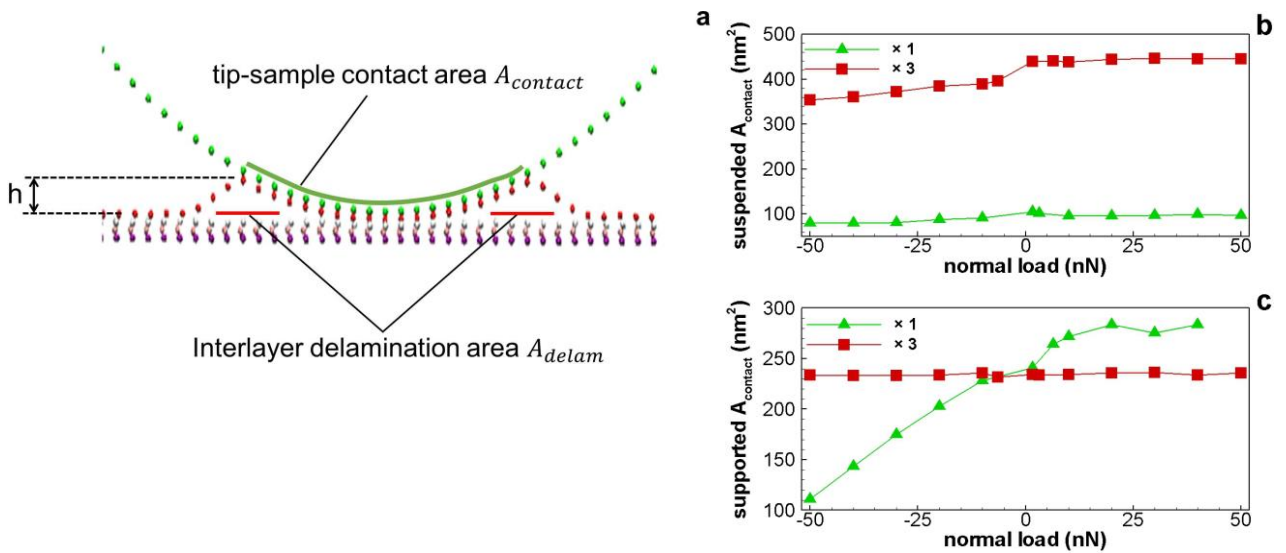

Figure 4. (a) Typical cross-section of the tip-sample contact in a CG simulation, outlining the contact quantities of interest. The bead colors outline the tip surface and separate layers in the stack. CG simulated $A_{contact}$ as a function of applied load for the tip-sample interaction strength equal to that of the interlayer interaction (×1) and a tip-sample interaction strength increased by a factor of three (×3), calculated for (b) the suspended and (c) the supported CG model.

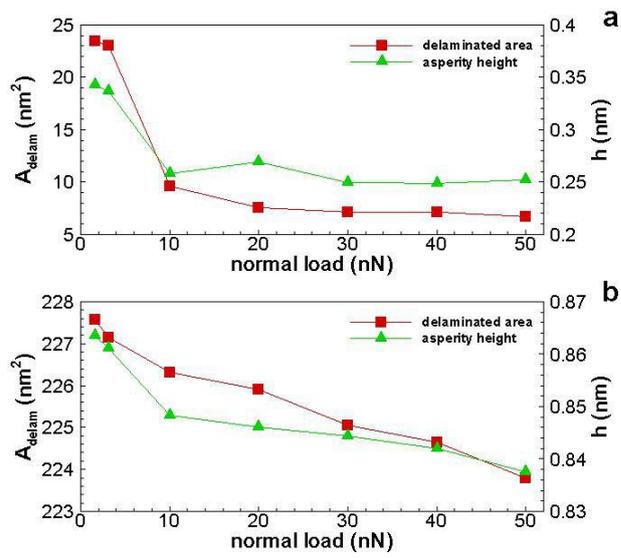

Figure 5. CG results of delaminated area $A_{delam}$ and asperity height $h$ in a supported system as a function of applied positive load for (a) a tip-sample interaction strength equal to that of the interlayer interaction, and (b) a tip-sample interaction strength increased by a factor of three.

**Supporting information for the paper:**
*Effects of surface compliance and relaxation on the frictional properties of lamellar materials*
by A. Smolyanitsky, S. Zhu, Z. Deng, T. Li, and R. J. Cannara

*Coarse-grained simulation model*
We adopt a coarse grained (CG) simulation scheme recently developed by Zhu *et al*. (Ref 23 of main text), which is recapped below and is similar to the model presented in [33]. In the CG scheme, the CG beads are organized in a hexagonal lattice (Fig. S1 (a)) for the simulation of a large sheet of monolayer graphene. The beads are treated as triangular particles, with the normal direction of the triangle perpendicular to the sample surface. The triangle-triangle center distance (CG bead bond length) is 0.568 nm. Effectively one bead represents 16 carbon atoms and an effective area of 0.419 nm$^2$. The van der Waals (vdW) interaction between two triangular particles is modeled by Lennard-Jones (LJ) interactions. The bonded energy terms consist of a two-body bond energy and three-body angle energy as $U_{bonding}(r_{ij}, \theta_{ijk}) = \sum \frac{1}{2} K_b (r_{ij} - r_0)^2 + \sum \frac{1}{2} K_\theta (cos\theta_{ijk} - cos\theta_0)^2$, where $r_{ij}$ is the distance between the $i^{th}$ and $j^{th}$ bonded CG beads with $r_0$ being its corresponding equilibrium distance; $\theta_{ijk}$ is the angle between the *i-j* bond and *j-k* bond with $\theta_0 = 120°$ being its corresponding equilibrium angle; $K_b$ and $K_\theta$ are bond force constant and angle force constant, respectively. The parameterizations of the bond force constant, angle force constant, and vdW parameters are described in detail in Ref. 23 of main text.

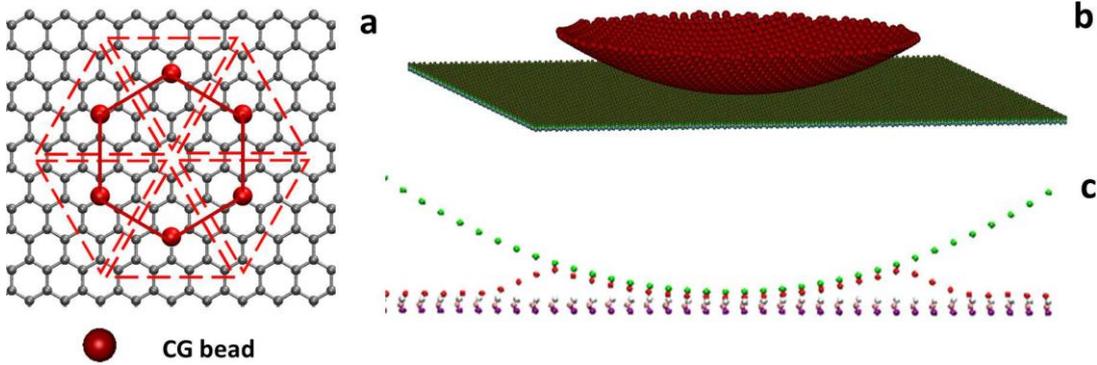

Figure S1. (a) CG interaction scheme (b) CG tip-sample configuration (c) A typical cross-section view of the geometry near the tip for enhanced tip-sample adhesion, where the bottom layer is fixed.

*Spectral energy of the measured stick-slip*
Because the stick slip process is not truly sinusoidal, a parametric study of $\delta$ was performed to ensure a reasonable convergence of $E(k_0)$ values, as shown in Fig. S2. In the main text, we used $\delta = 6.3$ nm$^{-1}$. It is noteworthy that integration of Eq. (2) (main text) from $2\pi/x_{max}$ (the shortest possible wave-vector along a trace of $x_{max}$ length) to a selected large wave-vector value effectively yields the average quadratic variation of $F(x)$ from the entire lateral trace dataset, including all sources of noise. Our calculations over a wide spectral range yielded data qualitatively similar to those obtained with the selected value of $\delta$, as shown in Fig. S2. All results presented in Fig. S2 and Fig. 2 (a) of main text were completely repeatable.
The calculation uncertainty in Eq. (2) of main text arises from the experimental error and, because it evaluates an integral of a quadratic function of the Fourier transform of $F(x)$, the relative uncertainty of calculating $E(k_0)$ is approximately twice the relative measurement error of $F(x)$. For the data presented in Fig. 2 (a) of main text, the average relative error is ≈ 6 %.

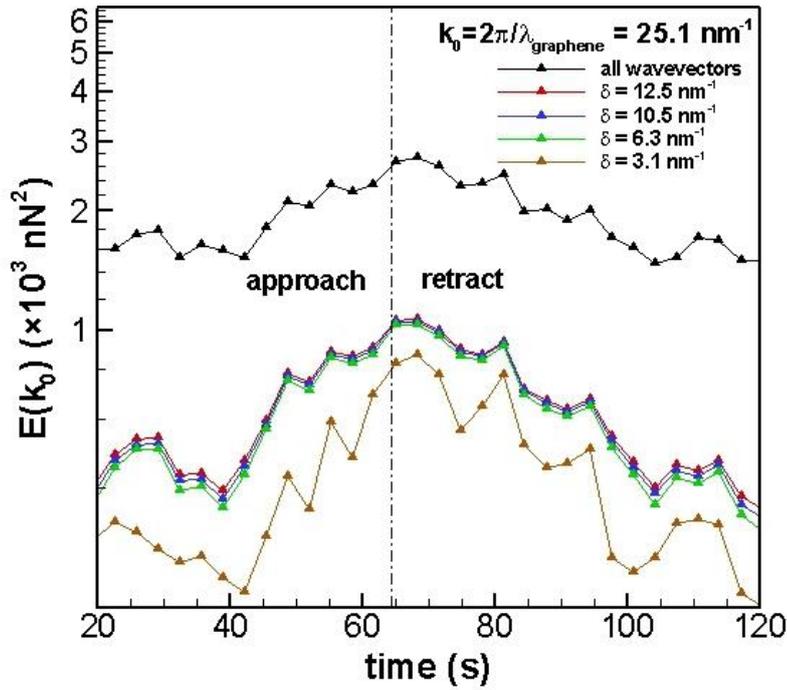

Figure S2. Parametric study of convergence of Eq. (2) (main text) for various values of δ, calculated for a freshly-cleaved graphite data set #1 (main text).

*Uncertainty in MD and CG calculations*
There is an inherent computational uncertainty associated with the quantities calculated in this work, arising from the choice of interatomic interactions, added thermal noise, as well as from the round-off integration errors. In the worst case, these uncertainties can reach the same order of magnitude as the measured quantities, *e.g.* resulting in systematic scaling of the amplitudes of the lateral force curves in Fig. 3 of main text. However, the qualitative trends within each presented curve, as well as comparisons of trends between the cases considered in the main text remain the same.

*Negative friction coefficient on aged $MoS_2$*
Variable-load experiments (similar to those on graphite) were performed on the molybdenum disulfide ($MoS_2$) surface. The friction-load relationship was obtained by ramping the AFM set point with force feedback on. Fig. S3 shows data acquired on the $MoS_2$ crystal aged overnight in laboratory air (≈ 25 % relative humidity at 21 °C). To test whether the observed friction-load relationship is reversible, we cycled the load setpoint in the AFM. Fig. S3a plots both raw deflection (blue line) and friction (red circles) signals from the experiment, from which it is apparent that the friction-load curves are reproducible. This is further affirmed by directly plotting friction as a function of load, as shown in Fig. S3b. Figs. S3a and S3b show that friction increases with load during the first tip approach andsubsequently retraces itself with repeated cycling of the load. These data support the notion that a reversible partial exfoliation is occurring also for aged $MoS_2$. The coefficient of friction ($\alpha$) is plotted as a function of pull-off force (exposure time) in Fig. S3c, demonstrating that $\alpha$ can be negative after exposure of $MoS_2$ to laboratory air. We note that the accelerated time frame for the transition to occur, as compared with that observed for graphite, posed a significant challenge to the $MoS_2$ measurements. Further work to control surface chemistry (transition time), would be required to produce similar data sets for $MoS_2$ as for graphite, and, for example, compare work of adhesion values with interlayer binding energies.

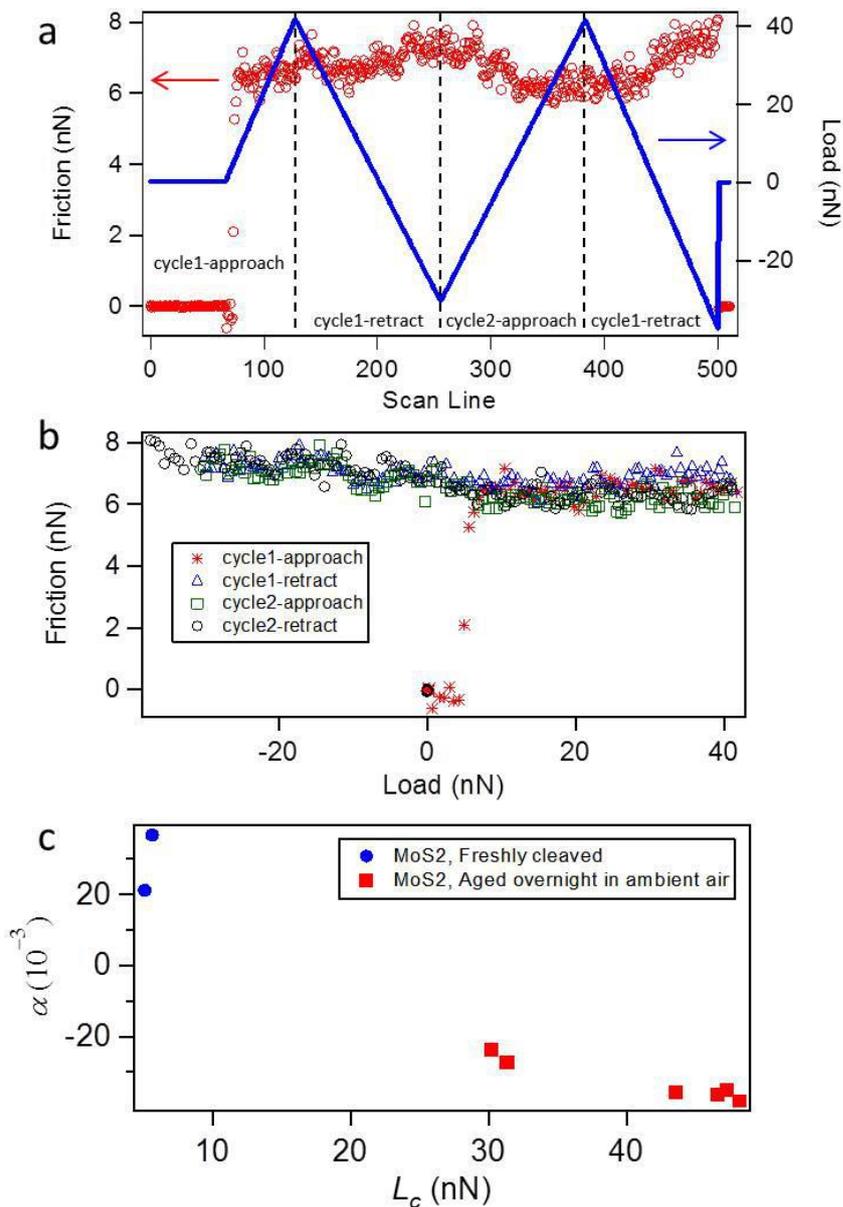

Figure S3. Reversible friction-load curves acquired with the 75 nm-radius ultrananocrystalline diamond probe in [7] on the surface of $MoS_2$. (a) Friction force and normal load plotted as the function of scanning line (time). (b) The same data in A plotted as a function of load instead of time. (c) Friction coefficient $α$ vs. adhesive force $L_C$ for $MoS_2$ freshly-cleaved and aged overnight in laboratory air.